\documentclass[]{spie}  %>>> use for US letter paper
\usepackage{amsmath,amsfonts,amssymb,float}
\usepackage{graphicx}
\usepackage[colorlinks=true, allcolors=blue]{hyperref}
\usepackage{gensymb}
\usepackage{geometry}

\title{Establishing Quantum-Secured Channels in Large-Scale Optical Networks}

\author[1]{Farzam Toudeh-Fallah}
\affil[1]{Quantum Communication and Photonic Systems, Ciena Corporation, 7035 Ridge Road, Hanover, MD 21076 USA}

\begin{document}

\newgeometry{top=0.5in,bottom=1.25in, left=0.88in, right=0.88in}

\begin{center}
    An extended version of this article is published under the title “Long-distance quantum-secured optical channels in operational environments,” at the Proceedings of SPIE, Volume 13148, Quantum Communications and Quantum Imaging XXII; 1314803 (2024)\\
    DOI: http://dx.doi.org/10.1117/12.3025871 [dx.doi.org]
\end{center}

{\let\newpage\relax\maketitle}

\begin{abstract}
Quantum-secured optical channels based on Quantum Key Distribution technology have generated a significant global interest. Although the maturity level of the short distance (less than 100 km) quantum-secured channels is at a deployment level, instituting such channels over long distance faces technological challenges, which is the subject of a world-wide research. In this article an industry perspective on establishing quantum-secured channels in large-scale optical networks in operational environments will be discussed, including the vision, requirements, and technical analysis of different approaches for establishing such channels.  
\end{abstract}


\section{INTRODUCTION}
\label{sec:intro}  % \label{} allows reference to this section

Quantum-secured optical channels based on Quantum Key Distribution (QKD) technology have generated a significant global interest towards achieving unconditional security \cite{Shore,Lo,Mayers}. QKD utilizes two channels (quantum and classical) to generate symmetric secret keys on both communication sites \cite{Bennett}. Quantum-secured optical channels are then established by utilizing these keys to encrypt the optical data channels. Although the maturity level of the short distance (less than 100 km) quantum-secured channels is at a deployment level \cite{Sandle}, instituting such channels in large-scale networks in which the nodes might be thousands of kilometers apart faces technological challenges and is the subject of world-wide research. 
As a consequence of the No-Cloning principle in quantum mechanics, optical amplifiers cannot be used for establishing long-distance quantum channels \cite{Muralidhara}. As such, two distinct approaches are considered by researchers to overcome this problem: terrestrial and satellite-based. In the terrestrial approach, the goal is to establish the long-distance quantum channel on a fiber optic link using quantum repeaters, while in satellite-based methodology the quantum channel is established between the satellite and ground stations. In this article, an industry perspective on the terrestrial approach with the intent of fostering pragmatic solutions towards deploying quantum-secured optical channels in large-scale operational networks will be discussed.

\section{VISION, APPROACH AND REQUIREMENTS}
\label{sec:vision}
A terrestrial approach must be pragmatic and adaptable towards deployment in the real-world operational environments. As such, the following requirements should be met in developing any terrestrial solution:

1) Fiber is an expensive commodity and as a result, dedicating fibers to quantum channels is not a pragmatic approach. Therefore, in practical deployments, quantum and optical data channels should be multiplexed on the same fiber. In addition, all degradation factors representative of the deployed fibers in operational networks must be taken into account in the performance analysis of the quantum channel established in an optical fiber.  

2) Quantum repeaters will be installed in the ILA (In-Line Amplifier) sites (also known as fiber huts), which are already deployed on average every 80 to 100 km along the fiber routs to amplify the signals in the optical networks. Therefore, quantum repeaters must be able to cover at least distances up to 100 km between them.

3) Quantum repeaters should be able to be deployed in the current ILA sites without the need for any additional sites.

\newgeometry{top=1in,bottom=1.25in, left=0.88in, right=0.88in}

4) The ILA sites mentioned above are very limited in available space and the power they can supply to the installed equipment and cannot provide any cryogenic environment. In addition to meeting the stringent requirements regarding space and power in ILA sites, quantum repeaters must be able to tolerate at least 40℃  ambient temperature without any cryogenic facility.

\begin{figure} [ht]
   \begin{center}
   \begin{tabular}{c} %% tabular useful for creating an array of images 
   \includegraphics[width=0.9\textwidth]{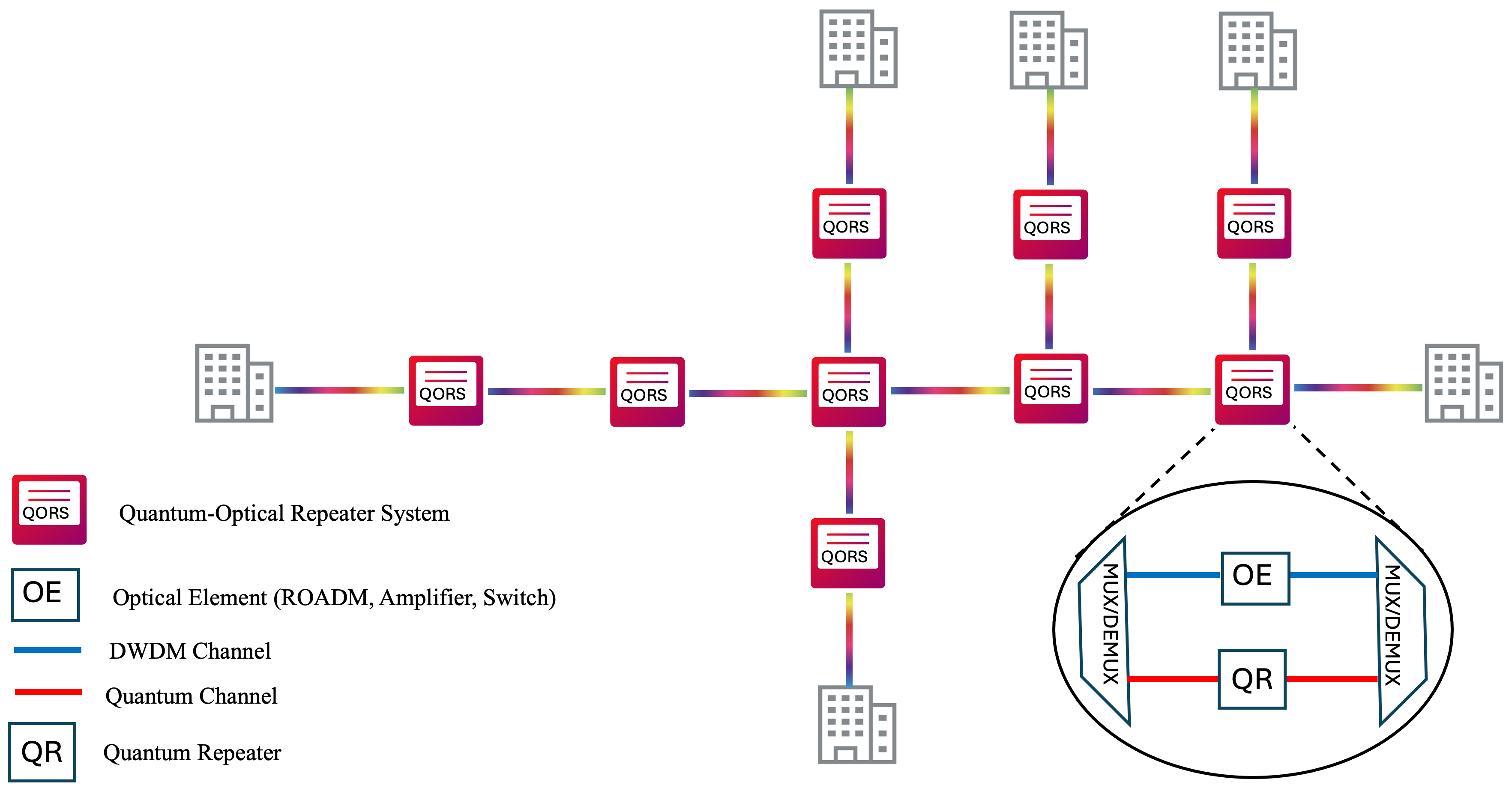}
   \end{tabular}
   \end{center}
   \caption[example] 
%>>>> use \label inside caption to get Fig. number with \ref{}
   { \label{fig:QuantumOptNetwork} 
Large-Scale Quantum-Optical Network}
   \end{figure}

Based on the above-mentioned requirements, Fig.~\ref{fig:QuantumOptNetwork} depicts how we envision the future deployment of terrestrial long-distance quantum-secured optical channels for operational environments in a large-scale quantum-optical network. In this model, the long-distance sites are connected via fiber optic links, in which DWDM optical data channels (C and L bands) and quantum channels (O band) are multiplexed on the same fiber (coexistence mode). The long distance between the remote sites is divided into several spans, with the span lengths defined in the requirement 2 mentioned above. These spans are connected to each other via devices that we designate as Quantum-Optical Repeater System (QORS) comprising optical elements and quantum repeaters. As depicted in Fig.~\ref{fig:QuantumOptNetwork}, optical and quantum channels are de-multiplexed at each QORS. The optical data channels are then routed to an optical element consisting of an optical amplifier, ROADM, switch or any other required optical element. The quantum channels are routed to the quantum repeater. After this operation, the optical and quantum channels are re-multiplexed and propagated to the next QORS over the fiber.

Optical elements are very well developed and installed on deployed networks. However, currently, quantum repeaters are still under intense world-wide research. Therefore, the focus of this article will be on the quantum channel and quantum repeater models and their applicability towards deployment in the real-world environments. 
  
\begin{figure} [ht]
   \begin{center}
   \begin{tabular}{c} %% tabular useful for creating an array of images 
   \includegraphics[width=0.9\textwidth]{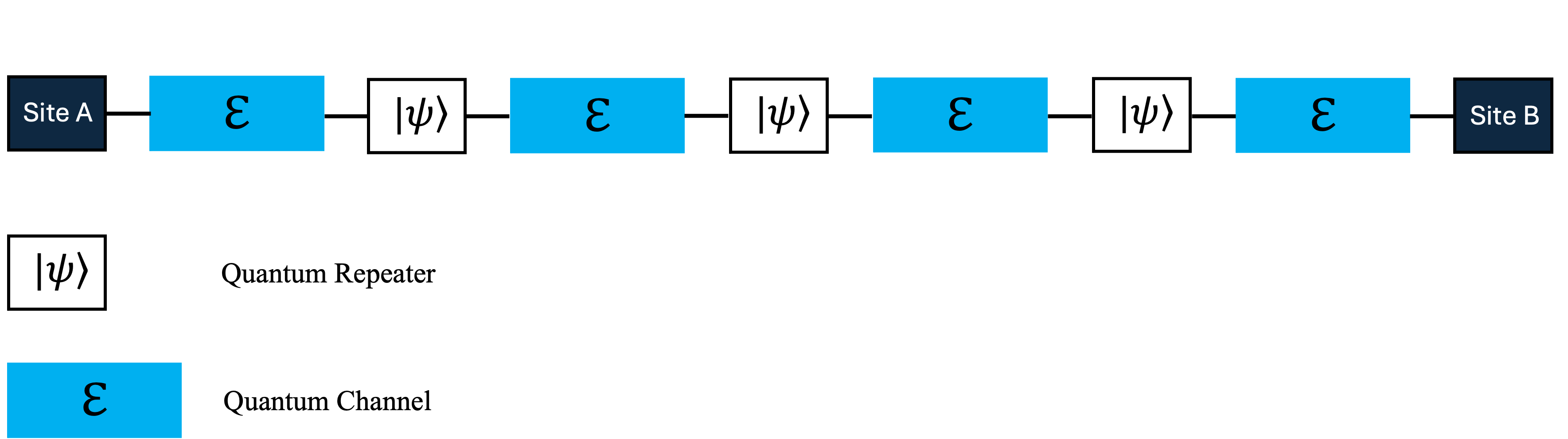}
   \end{tabular}
   \end{center}
   \caption[example] 
%>>>> use \label inside caption to get Fig. number with \ref{}
   { \label{fig:QuantumCommLink} 
Generic model for a long-distance quantum communication link}
   \end{figure}

Fig.~\ref{fig:QuantumCommLink} depicts a generic model for a long-distance quantum communication link, which consists of two major elements: quantum channel and quantum repeater.
In the following sections, the requirements for a quantum channel model representing the real-world environments and two of the most prominent approaches taken by researchers towards the development of quantum repeaters will be discussed.

\section{\uppercase{Requirements for a quantum channel model representing the real-world environments}}
   
Using the Stinespring Dilation model, the quantum channel, defined as a Completely Positive Trace Preserving map, is represented as a Unitary transformation of the joint system-environment state, as follows\cite{Stinespring,Nielsen}:

\begin{equation}
\label{eq:eq1}
\mathcal{E}( \mathcal{\rho}_{in}) = tr_E [U ( \rho_{in} \otimes \rho_E ) U^\dag ] \, ,
\end{equation}

where $\rho_{in}$ and $\rho_E$ represent the density operators of the input state and the environment, respectively, while $\mathcal{E}( \mathcal{\rho}_{in})$ represents the density operator of the output state. In equation \ref{eq:eq1}, defining U plays a key role in representing the nature of the quantum channel. A Gaussian quantum channel as a basis model for optical fibers, is represented by a Quadradic Bosonic Hamiltonian of the form \cite{Flynn,Zhang, Eisert}:

\begin{equation}
\label{eq:eq2}
U = e^{-i \hat{H}t}\, 
\end{equation}

\begin{equation}
\label{eq:eq3}
\hat{H} = \sum_{i,j=1}^{N} \big[ K_{ij} \hat{a}_i^\dag \hat {a}_j + \frac{1}{2} ( \Delta_{ij} \hat{a}_i^\dag \hat{a}_j^\dag + \Delta_{ij}^* \hat{a}_i \hat{a}_j )\big]\, 
\end{equation}

In equation \ref{eq:eq3} $\hat{a_i}$ and $\hat{a}_i^\dag$ are Bosonic annihilation and creation operators for mode i, respectively. Based on this approach, one can then work on deriving the operator-sum model for the specific channel\cite{Nielsen}:

\begin{equation}
\label{eq:eq4}
\mathcal{E} (\rho_{in}) = \sum_i K_i \rho_{in} K_i^\dag \quad ; \quad \sum_i K_i^\dag K_i = I \, 
\end{equation}
where $K_i$ represents the Kraus operator of the channel under investigation. Although there have been efforts to model the quantum channel and the associated Kraus operators for the real-world optical fibers, this medium has been mostly considered as purely represented by loss. However, there are several other degradation factors that must be considered in order to obtain the correct operator-sum model for the real-world environment representing an optical fiber. These include degradation factors such as State of Polarization (SOP) fluctuations, nonlinear effects such as Four Wave Mixing, Raman Scattering and dispersion. As an example, collected field data has indicated that SOP fluctuations, which can exceed 5 Mrad/s\cite{Charlton} can play a major role as a degradation factor in deployed optical fibers. Therefore, in order to model the quantum channel in real-world environments, in addition to loss, these models should take into account all these degradation factors, as well. 

\section{\uppercase{Quantum repeater models}}
Due to their role as the fundamental building block of quantum networks and long-distance quantum communication, quantum repeaters are currently the subject of intense world-wide research. Although there have been different approaches towards developing quantum repeaters, in this article two of the most prominent approaches and their applicability towards deployment in operational networks will be discussed.

\subsection{Entanglement Distribution}
\begin{figure}[H]
   \begin{center}
   \begin{tabular}{c} %% tabular useful for creating an array of images 
   \includegraphics[width=0.95\textwidth]{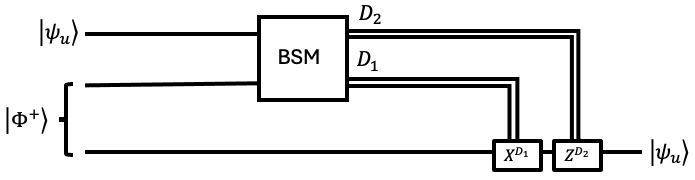}
   \end{tabular}
   \end{center}
   \caption[example] 
%>>>> use \label inside caption to get Fig. number with \ref{}
   { \label{fig:Teleportation} 
Quantum Teleportation used for transferring the quantum states between sites A and B}
   \end{figure}

In this approach, first each quantum repeater in Fig.~\ref{fig:QuantumCommLink} establishes entanglement with its neighbors. Then through 
\pagebreak
an iterative process known as Entanglement Swapping\cite{Gyongyosi}, entanglement between the two long-distance sites A and B will be established. At this point, Quantum Key Distribution could be conducted either by transferring the quantum states between the two sites using the well-known Quantum Teleportation process\cite{Nielsen} shown in Fig.~\ref{fig:Teleportation}, or by utilizing entanglement-based QKD protocols such as E91 or BBM92.

It should be mentioned that due to the elaborate process used to establish quantum entanglement between the two long-distance sites, which also includes two-way classical communications, one of the major requirements for this methodology is development of quantum memories with relatively long-coherence time to hold the quantum states intact during this process\cite{Knaut,Lipka}. Quantum memories currently under development in research labs require cryogenic environments. However, as pointed out before quantum repeaters will be ultimately deployed in the ILA sites, which impose very stringent set of requirements on space, power and ambient temperature of at least 40\degree C with no cryogenic environment. Although point-to-point entanglement distribution and swapping over optical fiber up to 100 km has been demonstrated\cite{Sun,Wengerowsky}, there are still major challenges to overcome for this approach towards deployment in operational networks. These challenges include but not limited to, achieving high coupling efficiency, high fidelity and acceptable entanglement generation rate between the two long-distance sites.

\subsection{One-Way Quantum Repeaters}
In this approach, Quantum Error Correction (QEC) methods are used by quantum repeaters to correct for both loss and operational errors\cite{Muralidhara}. In order to do so, quantum states are encoded in photons and any errors that occur during the transfer of these quantum states over the propagation medium is corrected by the QEC methods at each quantum repeater as photons pass through them. Therefore, this methodology eliminates the need for quantum memories and two-way classical communications, hence the name One-Way quantum repeaters. Although an attractive approach that eliminates some of the major challenges facing the Entanglement Distribution and Swapping approach, such as the need for quantum memories and cryogenic environment, this approach suffers from a major setback for deployment in operational networks. As mentioned in section \ref{sec:vision}, one of the requirements of a practical quantum repeater for deployment in operational networks is the ability to cover distances up to 100 km between the two repeaters. However, One-Way quantum repeaters utilize Quantum Error Correction as their fundamental principle, which can only tolerate up to 50\% loss in the link\cite{Muralidhara,Azuma,Azuma-2}. This would result into a hard threshold of 3 dB loss in an optical fiber. For an NDSF fiber, which is the most widely deployed fiber type in operational networks, this requirement would translate into a maximum distance of 15 km in C band and 8.6 km in O band, which is far below the required 100 km spacing between the quantum repeaters. Therefore, based on this argument, One-Way quantum repeater, or any other approach that is limited by the 3 dB link loss threshold of QEC is unacceptable as a pragmatic approach for deployment in operational networks.

% References
\bibliography{report} % bibliography data in report.bib
\bibliographystyle{spiebib} % makes bibtex use spiebib.bst

\end{document}